\documentclass[a4paper,conference]{IEEEtran}
\IEEEoverridecommandlockouts
\usepackage{cite}
\usepackage{amsmath,amssymb,amsfonts}
\usepackage{algorithmic}
\usepackage{algorithm}
\usepackage{graphicx}
\usepackage{textcomp}
\usepackage{xcolor}
\usepackage[colorlinks, citecolor=blue]{hyperref} 
\usepackage{tabularx} 
\usepackage{multirow}
\usepackage{subfigure}
\usepackage{dblfloatfix} 
\usepackage{enumitem}
\usepackage{geometry}
\newcommand{\eg}{{\it e.g., }}
\newcommand{\etal}{{\it et~al., }}
\newcommand{\ie}{{\it i.e., }}
\def\BibTeX{{\rm B\kern-.05em{\sc i\kern-.025em b}\kern-.08em
    T\kern-.1667em\lower.7ex\hbox{E}\kern-.125emX}}

\newlength{\boxfigwidth}

\begin{document}

\title{Descriptive and Predictive Analysis of Aggregating Functions in Serverless Clouds: the Case of Video Streaming




\author{
    Shangrui Wu$^{+*}$, Chavit Denninnart$^+$, Xiangbo Li$^\tau$, Yang Wang$^*$, Mohsen Amini Salehi$^+$ \\

    $^+$High Performance Cloud Computing (HPCC) Lab,     University of Louisiana at Lafayette, Louisiana, USA\\
    \{chavit.denninnart1, amini\}@louisiana.edu \\

    $^*$Department of Computer Science, Southwest Petroleum University, Chengdu, China\\
    \{201822000378, wangyang\}@swpu.edu.cn \\
    
    $^\tau$Twitch.tv, California, USA\\
     xiangbol@twitch.tv
}
}
\maketitle

\begin{abstract}
 Serverless clouds allocate multiple tasks (\eg micro-services) from multiple users on a shared pool of computing resources. This enables serverless cloud providers to reduce their resource usage by transparently aggregate similar tasks of a certain context (\eg video processing) that share the whole or part of their computation. To this end, it is crucial to know the amount of time-saving achieved by aggregating the tasks. Lack of such knowledge can lead to uninformed merging and scheduling decisions that, in turn, can cause deadline violation of either the merged tasks or other following tasks. Accordingly, in this paper, we study the problem of estimating execution-time saving resulted from merging tasks with the example in the context of video processing. To learn the execution-time saving in different forms of merging, we first establish a set of benchmarking videos and examine a wide variety of video processing tasks---with and without merging in place. We observed that 
 although merging can save up to 44\% in the execution-time, the number of possible merging cases is intractable.
 Hence, in the second part, we leverage the benchmarking results and develop a method based on Gradient Boosting Decision Tree (GBDT) to estimate the time-saving for any given task merging case. 
 Experimental results show that the method can estimate the time-saving with the error rate of 0.04, measured based on Root Mean Square Error (RMSE).

\end{abstract}

\begin{IEEEkeywords}
Task Merging, Oversubscription, Serverless, Cloud Computing, Video Stream Processing, Gradient Boosting Decision Tree (GBDT).
\end{IEEEkeywords}
\section{Introduction}

\begin{figure}
    \centering
    \includegraphics[width=0.4\textwidth]{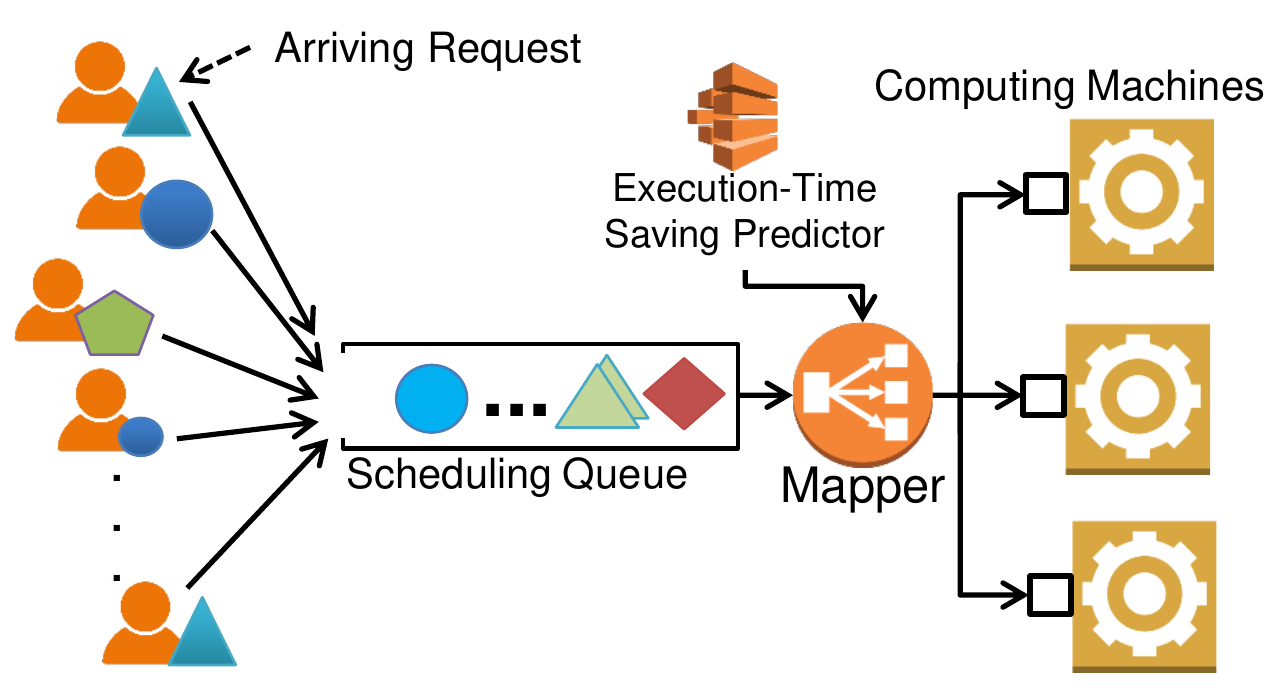}
    \caption{Tasks from multiple users are sent to a shared scheduling queue to be executed on computing resources. The execution-time saving predictor allows efficient use of computing machines. Geometries of different shapes, color, and size represent different (but can be similar) processing tasks.}
    \label{fig:intro_multiuser}
    \vspace{-10pt}
\end{figure}

In distributed computing systems, and particularly in the serverless cloud platforms, often multiple tasks (micro-services in the context of serverless clouds) are allocated on a set of shared resources \cite{lloyd2018serverless}. The resource sharing reduces the total resource consumption and subsequently achieves cost-efficiency.
In a serverless computing platform where resource sharing among multiple users is a norm, 
it is likely that multiple users independently request for an \textit{identical} or \textit{similar} task~\cite{denninnart2018leveraging}. For instance, in serverless platform specialized in video processing \cite{CVSSJournal}, two users can request to stream the same video with the same or different resolutions. Fig.~\ref{fig:intro_multiuser} shows a scenario where multiple users send their similar or identical service requests (tasks) to the system. Such tasks offer an opportunity to perform computational reuse for the requested service. The mapper (\ie scheduler) of the system is in charge of detecting these identical and/or similar tasks and initiating the reusing process.

Caching~\cite{andrade2019optimizing} is the established approach to enable reusing of \emph{identical} tasks. However, this approach cannot perform reusing for the executing and pending tasks. More importantly, the caching approach cannot achieve reusing for \emph{similar} tasks and uncacheable tasks such as those generated from live video streaming~\cite{vlsc}. A novel approach to achieve reusing for similar tasks is to aggregate them in the waiting and running states \cite{denninnart2018leveraging}. Aggregating (a.k.a. merging) of multiple tasks brings about multiple performance benefits, in terms of reducing the makespan time, and incurred cost requirement. 

\label{sec:intro}
However, the side-effect of task merging can be degrading the users' Quality of Service (QoS). In particular, rearranging and aggregating multiple small tasks create large tasks whose execution can potentially lead to deadline violation of either the merged task or other pending tasks scheduled behind it.

To avoid the side-effect of task merging and deadline violation, informed merging decisions should be made. Specifically, the mapper needs to know how much saving can be accomplished by merging two or more tasks and then, the merging is carried out, only if it is worthwhile. However, to date, a little attention has been paid in the literature to profile the execution-time of the merged tasks and understand their behavior. 
The challenge in profiling the task merging is that the number of possible combinations (\ie merging cases) is interactable and it is not feasible to examine and understand the behavior of all possible cases. Therefore, a method that can predict the execution-time of the merged task is required.

Accordingly, in this research, we \emph{first} strategically benchmark a variety of merging cases to understand the influential factors on merging effectiveness. 
Then, in the \emph{second} part, we develop a method (shown as Execution-Time Saving Predictor in Fig.~\ref{fig:intro_multiuser}) to estimate the execution-time saving resulted from merging any two or more given tasks. The proposed method operates based on a machine learning model that is trained using our observations in the first part.

Our motivational scenario is a serverless platform that is specialized in video processing (particularly, video transcoding \cite{CVSSJournal}) services. This platform processes video contents and formats them based on the viewers' display devices, internet bandwidth, and personal preferences \cite{icfec19vaughan,hcwali20}. The reason we concentrate on video processing is the increasing prevalence of video streaming in the Internet. Currently, video streaming constitutes more than 75\% of the Internet traffic \cite{tpds18perfor}. As such, even a minor improvement in video processing can play a significant role in reducing the cost and energy consumption on a global scale. In this context, we provide a benchmark of video segments and a set of tasks operating on those segments. We perform a descriptive analysis to understand the merging behavior for different number of merged tasks with various parameters. Then, we leverage the descriptive analysis and develop a method, based on Gradient Boosting Decision Tree (GBDT)~\cite{friedman2002stochastic}, to predict the execution-time saving of unforeseen merging cases. Although this initial study is focusing on video processing tasks. The methodology and the prediction model can be adapted to other contexts too.
In summary, the \textbf{key contributions} of this research study are as follows:
\begin{itemize}
\item  We collect and benchmark a video processing dataset that includes the execution-time of various video processing operations with and without task merging.
\item We provide a descriptive analyze of the influential factors on the execution-time saving of merged tasks.
\item We develop a method to predict the execution-time saving from merging any set of given tasks.
\end{itemize}

The rest of the paper is organized as follows: In Section ~\ref{sec:relwk}, we lay out background and related works to enhance video transcoding efficiency. 
Section~\ref{sec:exp} details the setup of the task merging experiments and examines the implications of the results.
Leveraging the obtained data, we propose and train a GBDT-based prediction model in Section~\ref{sec:learning}. 
Then in Section~\ref{sec:eval}, we optimize our prediction model and test the model prediction accuracy.
Finally, we conclude the paper and future work in Section~\ref{sec:conclsn}.

\section{Background and Related Works}
\label{sec:relwk}

\subsection{On-demand Video Processing}
Traditionally, video segments for video streaming are pre-processed and cached in multiple versions to fit the various device and user requirements. However, the pre-processing approach is cost-prohibitive and is not applicable for live streaming. On-demand video processing can overcome these limitations by processing each video to the user's exact specification upon request \cite{tpds18perfor}.

\begin{figure}[]
\centering
\includegraphics[width=0.4\textwidth]{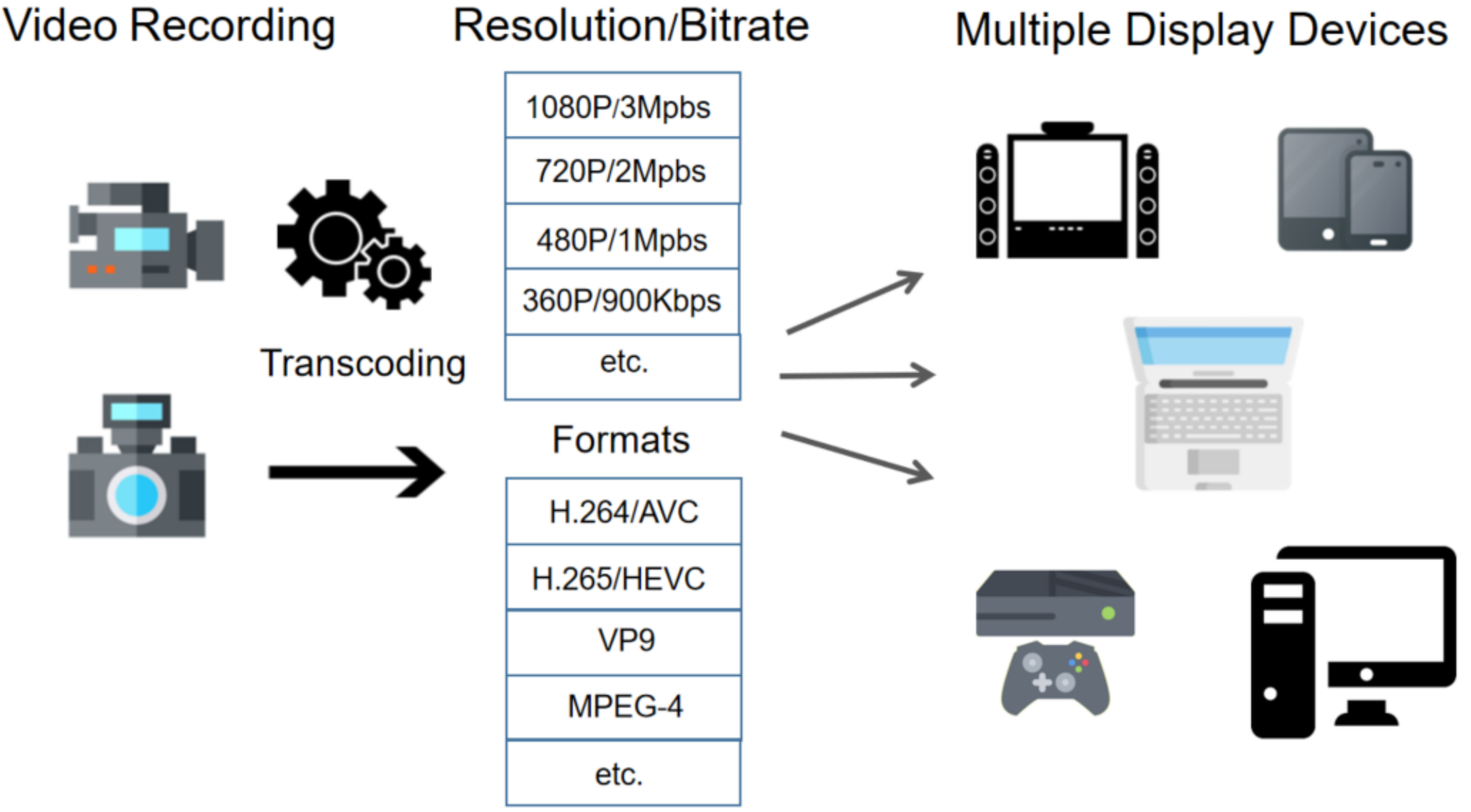}

\caption{A bird-eye view of video transcoding process flow. Videos are recorded at a very high bit-rate then transcoded to various compression standards to fit requirements of the network and display devices. }
\vspace{-12pt}
\label{fig:overview}
\end{figure}

In our prior work~\cite{denninnart2018leveraging}, we proposed an on-demand video streaming system based on a serverless cloud. 
In this system, video transcoding services (\eg altering codec, resolution, frame-rate, and bit-rate) transform the format of a source video to fit the viewer's device and bandwidth requirements. Fig.~\ref{fig:overview} shows a bird-eye view of recording videos in a certain format and then transcoding them before streaming to end-users with diverse display devices. 
Such a system frames the scope of this study.

\subsection{Detecting Different Types of Task Merging}

The nature of serverless cloud computing paradigm is to hide the resource provisioning and allocation decisions from the viewers' perspective \cite{denninnart2019improving}. This allows the cloud providers to handle the task scheduling and gain resource efficiency via aggregating viewers' tasks and avoiding redundant processing.
Tasks or services can be merged on the basis of some common properties, such as the same input data and/or the same operational process. The more properties the tasks have in common, the more potential exists to save in computing via merging the tasks together. In our prior study~\cite{denninnart2018leveraging}, we developed a method, with constant time complexity, to detect similarity between tasks by checking the hash signature of an arriving task against tables containing hash signatures of existing tasks. 
We categorize the task similarity levels of video tasks in three discrete types, namely \textit{Task level}, \textit{Data-Operation level}, and \textit{Data-only level}. Note that this categorization is arbitrary and can be categorized differently in other contexts.

Task level similarity indicates that the merging parties share all the relevant parameters for video transcoding. Therefore the task merging results in 100\% saving on the 2$^{nd}$ instance of the task by piggybacking on the first one. 
This is an evident type of reusing and we exclude it from our study.

Data-Operation level similarity is when the tasks are performing the same operation on the same video segment with different parameters. For instance, when two users request the same video at two different bit-rates. The video segment fetching (from the repository), decoding, and transcoding function loading can be merged. Only the bit-rate changing operation and final video encoding are performed separately. 
The merged task's execution-time is shorter than the sum of the time required to perform each task separately.

Data-Only level similarity is when the tasks are performing multiple different operations on the same video segment. 
In this type of merging in video processing example, only video fetching segment fetching and potentially decoding part can be shared while all other steps are proceed separately. 

While we have a rough idea of potential resource-saving in each form of merging, the exact magnitude of resource-saving is unknown and needs to be investigated in this study.

\subsection{Prior Studies on Benchmarking Video Processing}
Most prior studies on performance benchmarking and modeling in video transcoding focus on the performance of each video transcoding operation rather than the result of merging multiple requests. Here are some notable contributions. 

Netflix \cite{li2016toward} publishes a dataset to enrich the state-of-art video source for testing video quality metrics, the dataset contains 34 video clips from popular shows and movies, which embody multiple characteristics of video contents.
Furthermore, HD VideoBench \cite{perazzi2016benchmark} also provides some high definition digital videos for benchmarking. Those videos are encoded with MPEG-2, MPEG-4, and H.264. However, the selection of video content is limited (Blue sky, Pedestrian, Riverbed, and Rush hour) with three resolutions (1080P, 576P, and 720P).
Lottarini \etal~\cite{lottarini2018vbench} proposes Vbench which is a set of benchmark on video steaming workload captured from the cloud. From the collected video usage data, they algorithmically selects representative configurations with a more considerable variance.
They found that GPUs enabled cloud resources are the fastest configuration for high-quality live streaming scenarios.  

\section{Analysis of  Video Task Merging Operation}
\label{sec:exp}
\subsection{Video Benchmark Dataset}
We used 3,159 video segments to construct the benchmark dataset. The video segments are gathered from a set of 100 open-license videos in YouTube \cite{youtube}. To build a representative dataset, we assured that the chosen videos cover diverse content types with distinct motion patterns (\ie fast or slow pace) and various object categories. 

\begin{table}[htb]
\centering
\normalsize
\setlength\tabcolsep{1pt}
\scalebox{0.85}{
\begin{tabularx}{0.561\textwidth} { 
   >{\centering\arraybackslash}X 
  | >{\centering\arraybackslash}X 
  | >{\centering\arraybackslash}X 
  | >{\centering\arraybackslash}X 
  | >{\centering\arraybackslash}X
   >{\centering\arraybackslash}X}
  & \textbf{Codec} & \textbf{Frame-rate} & \textbf{Resolution} & \textbf{Container}  \\
 \hline
 \hline
 \textbf{\small{Standardized format}} & H.264 (High) & 30 fps & 1280$\times$ 720   & MPEG transport stream (TS)\\
\end{tabularx}}
\\
\vspace{5pt}
\caption{Standardized specifications for videos in the collected video benchmark dataset.}
\label{table:1}
\vspace{-20pt}
\end{table}%

To systematically analyze the evaluation results and eliminate the impact of different video formats that affect the execution-time, we split all the videos to two-second video segments with the standardized format detailed in Table \ref{table:1}. It is noteworthy that segmenting videos is a common practice in stream providers and the two-second is to comply with the MPEG transport streaming~\cite{fairhurst2005unidirectional,alzahrani2018impact} standard. We choose H.264 as the unified codec, because it is still the most common and widely compatible format for video streaming. 
We selected libx264 \cite{x264} as the encoders to change all the proposed video formats. The benchmark dataset contains 3,159 video segments that are publicly available\footnote{\url{https://bit.ly/3gKNijT}} 
for reproducibility purposes, 
with detailed description of the each video\footnote{\url{https://bit.ly/2YMIwwb}}.

\begin{figure*}

\centering

\subfigure[Bit-rate]{
    \centering
    \includegraphics[width=0.314\textwidth]{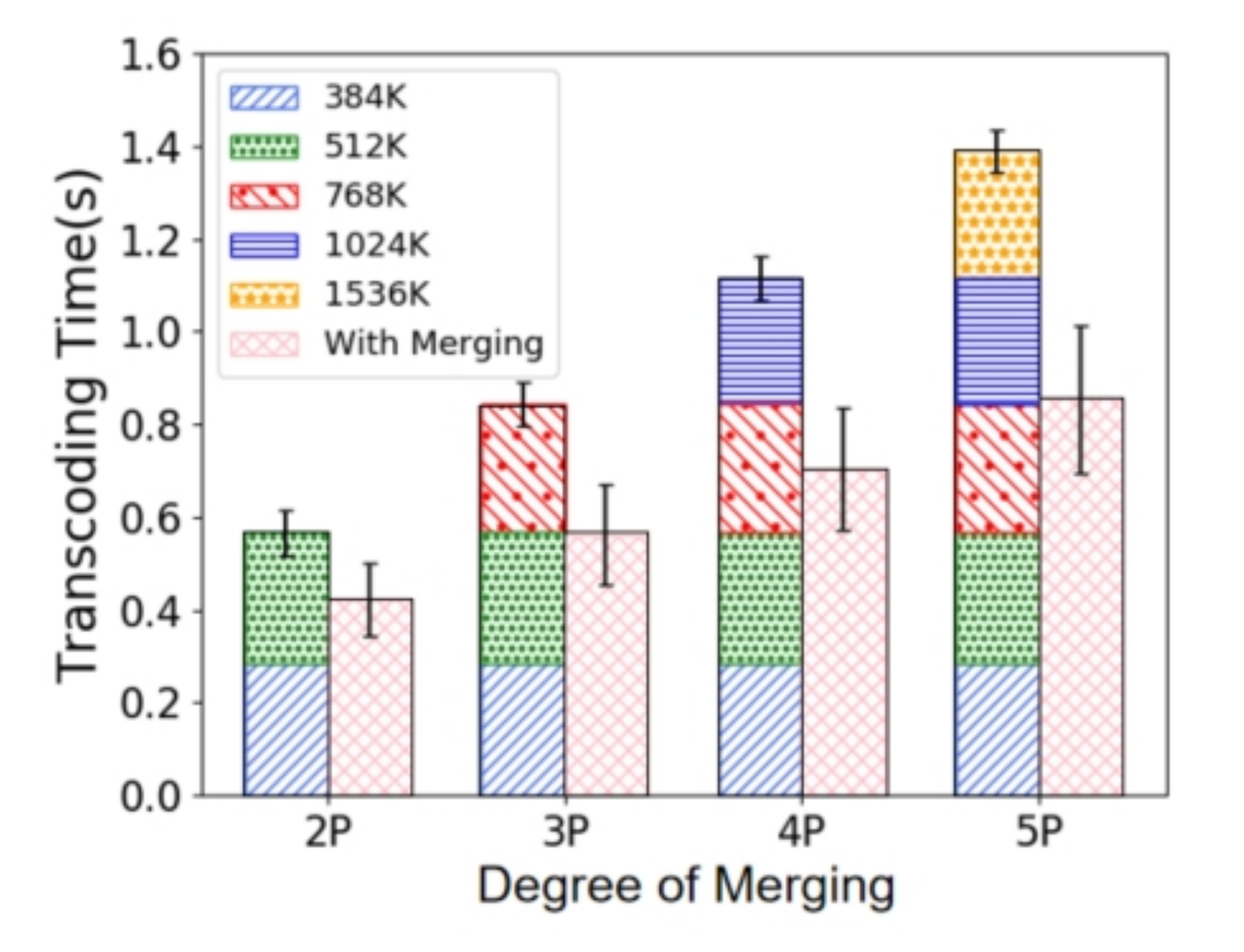}
    \label{fig:result-singlea}
}
\subfigure[Frame-rate]{
    \centering
    \includegraphics[width=0.307\textwidth]{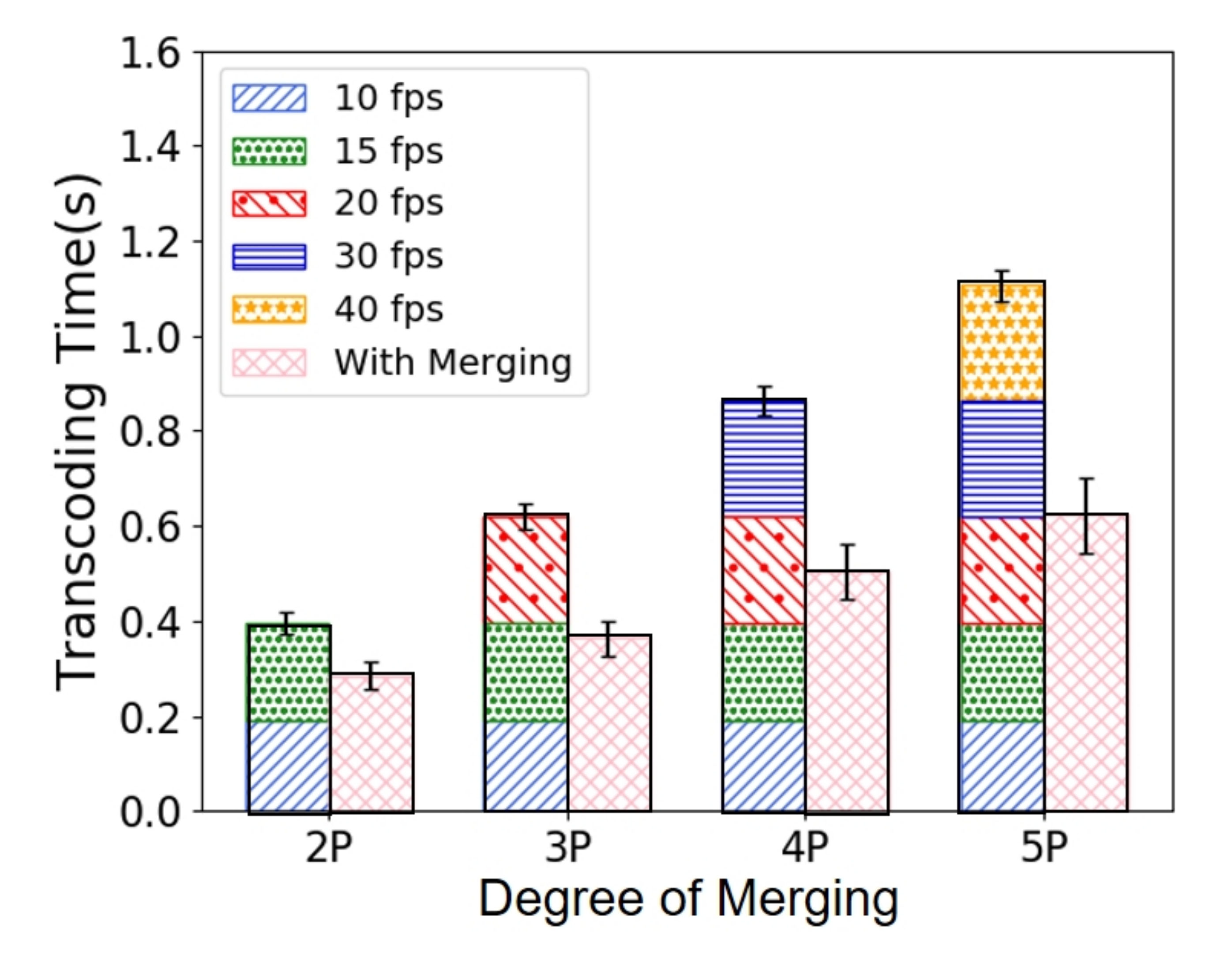}
    \label{fig:result-singleb}
}
\subfigure[Resolution]{
    \centering
    \includegraphics[width=0.295\textwidth]{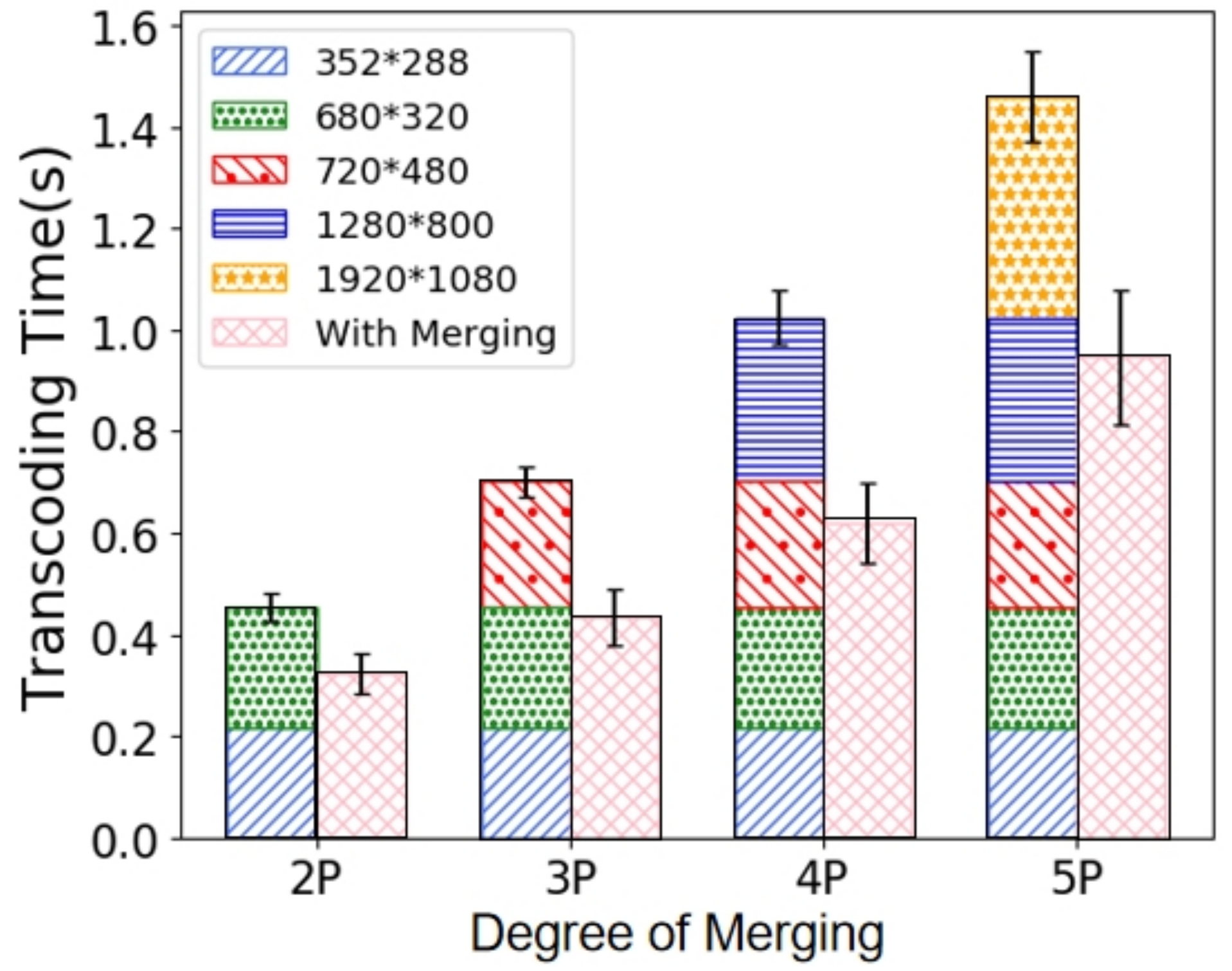}
    \label{fig:result-singlec}
}

\caption{Comparison of the total transcoding time (\ie makespan) (in seconds) to execute multiple tasks with two to five parameters (2P---5P in the horizontal axes) within the VIC group in two scenarios: executing individual tasks sequentially (without task merging) versus executing them as a merged task. Sub-figures (a), (b), and (c) represent transcoding time of bit-rate changing operation, frame-rate changing operation, and resolution changing operation, respectively.}
\label{fig:result-single}
\vspace{-10pt}
\end{figure*}

\subsection{Benchmarking Execution-Time of Video Transcoding Tasks}
\label{subsec:collection}
Based on the video segments of the collected dataset, we perform a set of benchmark services that consists of four primary video transcoding operations (tasks), namely changing bit-rate, frame-rate, resolution, and codec. Early evaluation of the collected execution-time revealed a remarkable variation in the execution-time of some task types. Specifically, we noticed that codec execution-time is far beyond the other three task types.
Accordingly, we categorize the tasks types into two groups: \emph{First} group is called Video Information Conversion (\emph{VIC}) that includes changing bit-rate, frame-rate, or resolution task types. Tasks of this group have a low variation in their execution-times, when processing different video segments on the same machine type. \emph{Second} group is Video Compression Conversion that only includes the codec task type (hence, we call it the Codec group). In contrast to the first group, the codec execution-time  (and subsequently its merge-saving) for different video segments varies remarkably even on the same machine.

\begin{table}[htb]
\normalsize
\centering
\setlength\tabcolsep{1pt}
\scalebox{1}{\begin{tabular}{c|c|c|c|l|l}

                                                                               \multicolumn{3}{c|}{\small{\textbf{\small{Video Information Conversion (VIC)}}}}                                                                    & \multicolumn{3}{c}{\multirow{2}{*}{\small{\textbf{Codec}}}} \\ \cline{1-3}
                                                                               \multicolumn{1}{c|}{\small{\textbf{Bit-rate}}} & \multicolumn{1}{c|}{\small{\textbf{Frame-rate}}} & \multicolumn{1}{c|}{\small{\textbf{Resolution}}} & \multicolumn{3}{c}{}                       \\ \hline \hline 
                                                                    
                                                                             384K                          & 10 fps                             & 352$\times$288                         & \multicolumn{3}{c}{\small{MPEG-4}}                 \\ 
                                                                              512K                          & 15 fps                             & 680$\times$320                         & \multicolumn{3}{c}{\small{H.265/HEVC}}             \\ 
                                                                              768K                          & 20 fps                             & 720$\times$480                         & \multicolumn{3}{c}{\small{VP9}}                    \\ 
                                                                              1024K                         & 30 fps                             & 1280$\times$800                        & \multicolumn{3}{c}{-}                       \\ 
                                                                              1536K                         & 40 fps                             & 1920$\times$1080                       & \multicolumn{3}{c}{-}                       
\end{tabular}}
\vspace{5pt}
\caption{The list of parameters employed to form various transcoding tasks. Each transcoding task changes only one specification of the videos in the standardized benchmark dataset. Accordingly, there are collectively 18 transcoding tasks: 5 for bit-rate changing, 5 for frame-rate changing, 5 for resolution changing, and 3 for codec changing.} 
\label{table:2}
\vspace{-10pt}
\end{table}

To limit the degree of freedom in execution-time, we configured each transcoding task to change only one specification of the videos in the benchmark dataset. The characteristics (parameters) of the evaluated transcoding tasks are listed in Table~\ref{table:2}. According to the table, there are 4 task types and collectively 18 transcoding tasks, including 5 different parameters in tasks changing bit-rate, 5 parameter for tasks changing frame-rate, 5 parameters in tasks that change resolution, and 3 parameters in tasks changing codec. 

To evaluate a variety of task merging cases, we compare the time difference between executing the 18 video transcoding tasks individually against executing them in various merged forms. 
Our preliminary evaluations showed that there is little gain in merging more than five tasks. In addition, we observed that it is unlikely to find more than five (similar, but not identical) mergeable tasks at any given moment in the system
~\cite{denninnart2018leveraging, Chavit2020Leveraging}.
As such, in the benchmarking, the maximum number of merged tasks (a.k.a. degree of merging) is limited to five. Even with this limitation, exhaustively examining all possible permutations of merging 18 tasks (in batches of 2, 3, 4, 5 tasks) collectively leads to $C(18,2)+C(18,3)+C(18,4)+C(18,5)$ cases, where $C(x,y)$ 
refers to $y$-combinations from a set of $x$ tasks. That entails 12,597 experiments per video segment.
As performing this many experiments is time prohibitive, we reduce the number of possible test cases to some highly representative merging cases for each video segment. Details of the conducted benchmarking is as follows:

\begin{enumerate}[label=(\Alph*)]
    \item We measured the execution-time of the 18 tasks on each one of the 3,159 video segments in the dataset individually. This means that, in this step, we collected 56,862 execution-times for individual tasks. 
    \item We measured the execution-time of merged tasks with the same operation and 2---5 various parameters. That is, each merged transcoding task is composed of one operation (\eg changing resolution) with two to five different parameters (\eg based on the possible values of resolution, mentioned in Table~\ref{table:2}). Then, to measure the magnitude of saving resulted by the task merging (henceforth, referred to as \emph{merge-saving}), the resulting execution-times are compared against execution-time of individual tasks, generated in Step (A).
   \item In our initial evaluations, we observed more consistent behavior in merge-saving of the VIC group, as opposed those mergings included codec. As such, our evaluations were focused on the merging cases with various operations within the VIC group. Each operation can have various parameters. For instance, consider video $A$ with bit-rate $b_1$, frame-rate $f_1$, and resolution $r_1$. We merge multiple transcoding tasks on $A$ to change: its resolution to $r_2$, its bit-rate to $b_2$ and its frame-rate to $f_2$ and $f_3$. Then to measure the magnitude of merge-saving, the resulting execution-times are compared against execution-time of individual transcoding time from (A).
    \item We benchmark and analyze execution-time of merged tasks with codec operation and operations from the VIC group. The process is similar to (C). However, each merged task is composed of one codec changing operation with one or more VIC class operations.
 \end{enumerate}

\subsection{Analyzing the Impact of Task Merging on Execution-Time}
\subsubsection{Evaluating the impact on the makespan time}
To understand the task merging performance behavior, we evaluate the total transcoding time (a.k.a. makespan) of the tasks in the VIC group under two scenarios: transcoding with and without merging. We consider merging of two to five parameters for bit-rate, frame-rate, and resolution separately---shown as $2P$ to $5P$ in the horizontal axes of Fig.~\ref{fig:result-single}. 
The difference between transcoding time when executing each task individually versus when the tasks are merged represents the merge-saving. 

We observe that, in all cases, there is an increasing trend in the merge-saving when the degree of merging is increased.
Interestingly, we observe that the ratio of merge-saving generally increases for the higher degrees of merging. The only exception is in Fig.~\ref{fig:result-singlec} (changing resolution) that by increasing the degree of merging from 4P to 5P, the merge-saving ratio is not increased. 
In general, we can conclude that all task merging with operations within the VIC group consistently and substantially save the execution-time.


\subsubsection{Evaluating the impact on execution-time saving}

\begin{figure}
    \centering
    \includegraphics[width=0.42\textwidth]{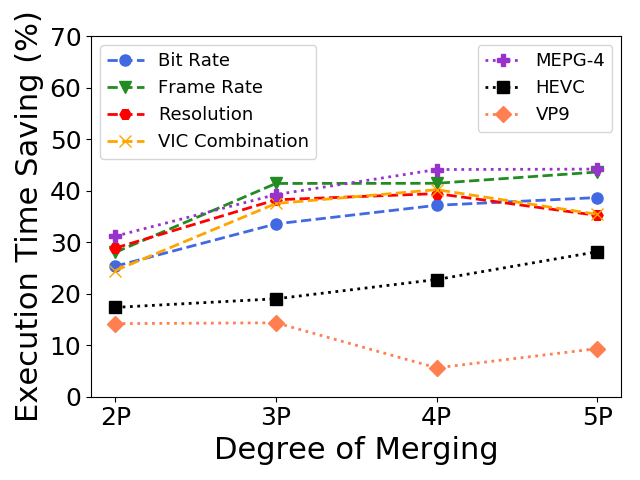}
    \caption{The result of merge-saving across varying numbers of the videos transcoding tasks. Figure (a) and (b) show the makespan saving when tasks merged within the VIC group and the makespan saving when codec transcoding tasks merged with VIC group, respectively.}
    \label{fig:result-sum}
    \vspace{-12pt}
\end{figure}

Changing the view to focus on execution-time saving percentage, Fig.~\ref{fig:result-sum} shows that, on average, when two tasks in the VIC group are merged ($2P$), the execution-time is saved by 26\%. 
The saving increases to 37\% when three tasks merged together. From there, the saving taper off to around 40\% for four and five tasks merging (4P and 5P).  We do not observe significant extra merge-savings after 5P. In addition, forming a large merged task complicates the scheduling and increase the potential side-effects (in the form of delaying) the completion of the large task itself or other pending tasks \cite{Chavit2020Leveraging}. This observation holds for the merged tasks compose of multiple different operations within VIC group (denoted as \textit{VIC Combination}).



For merged tasks that include codec changing operations, the results are far from consistent. Merge-saving of tasks that include MPEG-4 codec changing behave similarly to pure VIC group operations. Merge-savings of tasks with HEVC codec changing operation are consistently lower than any aforementioned cases for every degree of merging. 
The minimum saving is observed when the merged task includes VP9 codec changing operation. In which case, the saving is even reduced when the degree of merging increased from 3P to 4P. 


The results suggest that the significant gain in merging takes place in the first three tasks merging.
We can conclude that, to strike a balance between efficiency gain and potential side-effects of task merging, the system should target to form groups of about three tasks, rather than forming the biggest possible group of task merging. 
It is also worth mentioning that codec changing operations have a significantly (up to eight times) longer execution-time than VIC group operations. Merging a codec changing task to VIC group tasks does not necessarily offer a significant merge-saving, yet can jeopardizes the users' QoS. That is, merging a short task from the VIC group to a large task from the codec group can significantly delay the completion time of the short task and degrades its QoS (\eg in terms of missing the task's deadline). 


\section{Predicting the Execution-Time Saving of Task Merging}
\label{sec:learning}

\subsection{A Model to Predict Execution-Time Saving }
In the benchmarking process, 
we noticed that the number of cases that tasks can be merged in a system is interactable (see Section~\ref{subsec:collection}). That is, it is not feasible to pre-generate the knowledge of the merge-saving of all task types with all possible parameter values and for all video files. However, such a knowledge is crucial to decide about performing a task merging case \cite{Chavit2020Leveraging}.  
As such, our goal in this part is to leverage our findings in the benchmarking section and develop a machine learning model that can predict the merge-saving of any given set of mergeable tasks based on the task types and characteristics of the video segments.

In total, 81,327 data points, obtained from the benchmarking, were used to train the proposed model. For training and validating the model, we extracted metadata of the benchmark videos and transcoding configurations. A short sample of these metadata is shown in Table~\ref{table:3}.
As we can see in the table, for each video, we collected its essential static features, including duration, segment size, frame-rate (FR), width, and height (for the sake of better presentation, only few columns are shown in the table). 
Then, we concatenate the static features to the specification of merged task's transcoding configuration. 
The transcoding configuration includes the number of bit-rate changing (B), spatial resolution/frame-rate changing (S), resolution changing (R), and the type of codec changing included in the merged task. 
The output of the machine learning model is the merge-saving, \ie the percentage of improvement in execution-time upon merging several tasks versus not merging them. 

Since the three codec transcoding parameters behave significantly different, the codec operation parameters are marked separately in Table~\ref{table:3}, as MPEG4, VP9, and HEVC columns. 
In contrast, for the ones in the VIC group, we observed that their configurations (\ie parameter values) have little influence on the merge-saving, in compare with their degree of merging.  As such, for elements of the VIC group, we consider the number of operations (sub-tasks) in the merged task as opposed to the value of their parameters. Accordingly, the integer values in the B, S, and R columns represents the number of those operations included in the merged task. The main benefit of marking the table in this manner is to create a robust model that can infer the merge-saving even for unforeseen parameters. Arguably, if we bind the elements of VIC group to their parameter values in the training, then the model cannot efficiently predict the merge-saving of a merge request whose parameter values are out of the scope of the training dataset.

\begin{table}[htb]
\centering
\setlength\tabcolsep{0.5pt}
\begin{tabular}{c|c|c|c|c|c|c|c|c|c|c|c}
\textbf{\small{Dura-}} & \textbf{\small{Size}} &  \textbf{\small{FR}} & \textbf{\small{Width}} & \textbf{\small{Height}}  & \textbf{\small{B}}   & \textbf{\small{S}}   & \textbf{\small{R}}   & \textbf{\small{MP-}} & \textbf{\small{VP9}} & \textbf{\small{HEVC}} & \textbf{\small{Saving}} 
\\
\textbf{\small{tion (s)}} & \textbf{\small{(KB)}} &  &  &   &   &    & &\textbf{\small{EG-4}}  &  & &
\\ \hline
\hline

2.0 & 876  & 30  & 1280  & 720    & 1   & 0   & 0   & 1     & 0   & 0    & 33.60\% \\

2.0 & 1085 & 30  & 1280  & 720    & 1   & 2   & 1   & 0     & 0   & 0    & 39.17\% \\

2.0 & 1231 & 30  & 1280  & 720    & 1   & 1   & 1   & 0     & 1   & 0    & 20.22\%  \\

1.2 & 969  & 30  & 1280  & 720    & 0   & 0   & 1   & 0     & 1   & 0    & 27.89\% \\

2.0 & 864  & 30  & 1280  & 720    & 1   & 3   & 1   & 0     & 0   & 0    & 23.33\% \\

2.0 & 1091 & 30  & 1280  & 720    & 1   & 1   & 1   & 0     & 0   & 1    & 21.95\%  \\

0.9 & 347  & 30  & 1280  & 720    & 1   & 0   & 1   & 0     & 0   & 0    & 31.32\% \\

...      & ...  & ...       & ...   & ...    & ... & ... & ... & ...   & ... & ...  & ...\\
\end{tabular}
\vspace{4pt}
 \caption{A sample of the training dataset. Left side columns show static features of videos, such as duration, size, frame-rate (FR), and dimensions. B, S, and R columns
 represent bit-rates, frame-rate, and resolution changing operation sub-tasks in the particular merged task. Codec changing operation parameters are marked separately with one possible parameter per column (as MPEG-4, VP9, and HEVC.) The Saving column indicates the merge-saving caused by a particular task merging.}
\label{table:3}
\vspace{-10pt}
\end{table}

\subsection{Gradient Boosting Decision Tree (GBDT) to Predict the Execution-Time Saving}





Decision tree~\cite{vadim2018overview} is a known form of prediction model that functions based on a tree-based structure. Starting from the head node, the model performs a test on a feature at each one of its internal nodes. Ultimately, the traversal leads to a leaf node that includes the prediction~\cite{magerman1995statistical}. In particular, decision trees are proven to be appropriate for predicting numerical of unknown data~\cite{kotsiantis2013decision}. Because merge-saving prediction can be considered as a kind of numerical prediction problem, we choose decision trees to predict the saving. However, solutions based on a single decision tree are generally prone to the over-fitting problem~\cite{kotsiantis2013decision}. That means, the model is excessively attached to the training dataset such that, at the inference time, its prediction cannot cover slight variations in the input. 

Accordingly, to devise a prediction model that is robust against over-fitting, we utilize a optimal method of decision trees, known as Gradient Boosted Decision Trees (GBDT)~\cite{ friedman2002stochastic}. This is an iterative construct based on boosted ensemble of weak-learner decision trees. In fact, GBDT combine the multiple boosted weak-learners into a high accuracy and robust model. The boosting technique uses a process in which subsequent predictors learn from errors of the previous predictors. The objective of each iteration is to reduce the prediction error, which is calculated by a loss function~\cite{friedman2002stochastic}.



The pseudo-code, shown in Algorithm~\ref{fig:boosting}, elaborates on how the merge-saving prediction model is trained based on GBDT. On line 2 of the pseudo-code, a subset of the benchmark dataset, explained in Section \ref{sec:exp}, is generated and is used as the training dataset, denoted as $t$. We considered 80\% of the benchmarked dataset in $t$. The initial decision tree, denoted as $B_0(x)$, is created with random number and trained based on $t$ on line 3. On line 4, the main loop of the training model aims at creating one weak model based (decision tree) per iteration. 
Note that $x$ represents the input features of the merged task, as expressed in Table~\ref{table:3}.
In this step, there are various hyper-parameters that affect form of the decision tree being created. Notable hyper-parameters (among many others~\cite{kotsiantis2013decision}) that impact the accuracy of the prediction model are the learning rate (denoted as $L$), maximum depth of the individual regression estimators (denoted as $D$), the minimum number of samples required to split an internal node (denoted as $S$), and the minimum number of samples needed to be at a leaf node (denoted as $J$). In Sections~\ref{subsec:evalL}---\ref{subsec:evalSJ}, we elaborate on the appropriate values of these hyper-parameters such that the prediction accuracy of the merge-saving prediction model is maximize. 

\begin{algorithm}[h]
    \caption{Pseudo-code of the method to build the prediction model of the execution-time saving of a merged task.}
    \label{fig:boosting}
    \begin{algorithmic}[1]
        \REQUIRE
            The merge-saving benchmark dataset $T$, obtained from Section~\ref{sec:exp};
        \ENSURE
            Execution-time saving predictor $B_M(x)$;
        \STATE Let $M$ be the number of decision trees (and iterations)
        \STATE Create training dataset t, where $t \subset T$;
        \STATE Initialize decision tree $B_0(x)$ from $t$;
        \FOR{$m \leftarrow 1$ to $M$} 
            \STATE $r_{mi} \leftarrow$ Compute the prediction error of the $B_{m-1}(x)$;
            \STATE Utilize ($x_{i}$, $r_{mi}$) to fit a regression tree, calculating the fitted values for each terminal region;
            \STATE Update $B_m(x)$ based on the $B_{m-1}(x)$;
        \ENDFOR 
        \RETURN The merge-saving prediction model $B_M(x)$;
    \end{algorithmic}
\end{algorithm}

 \begin{figure*}
\centering

\subfigure[Number of Trees (M)]{
    \includegraphics[width=0.31\textwidth]{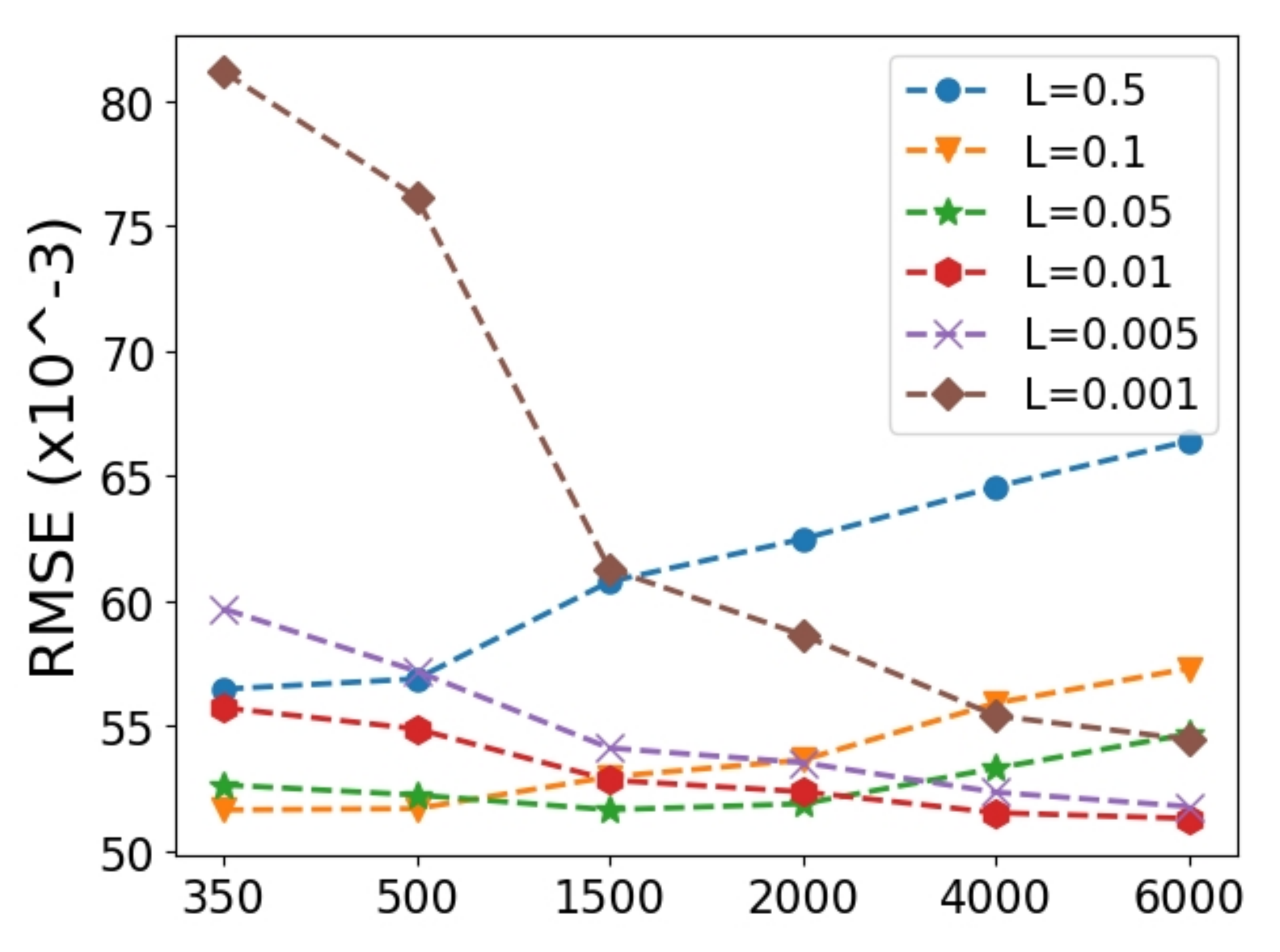}
    \label{fig:estimators}
}
\subfigure[Maximum Depth (D)]{
    \includegraphics[width=0.31\textwidth]{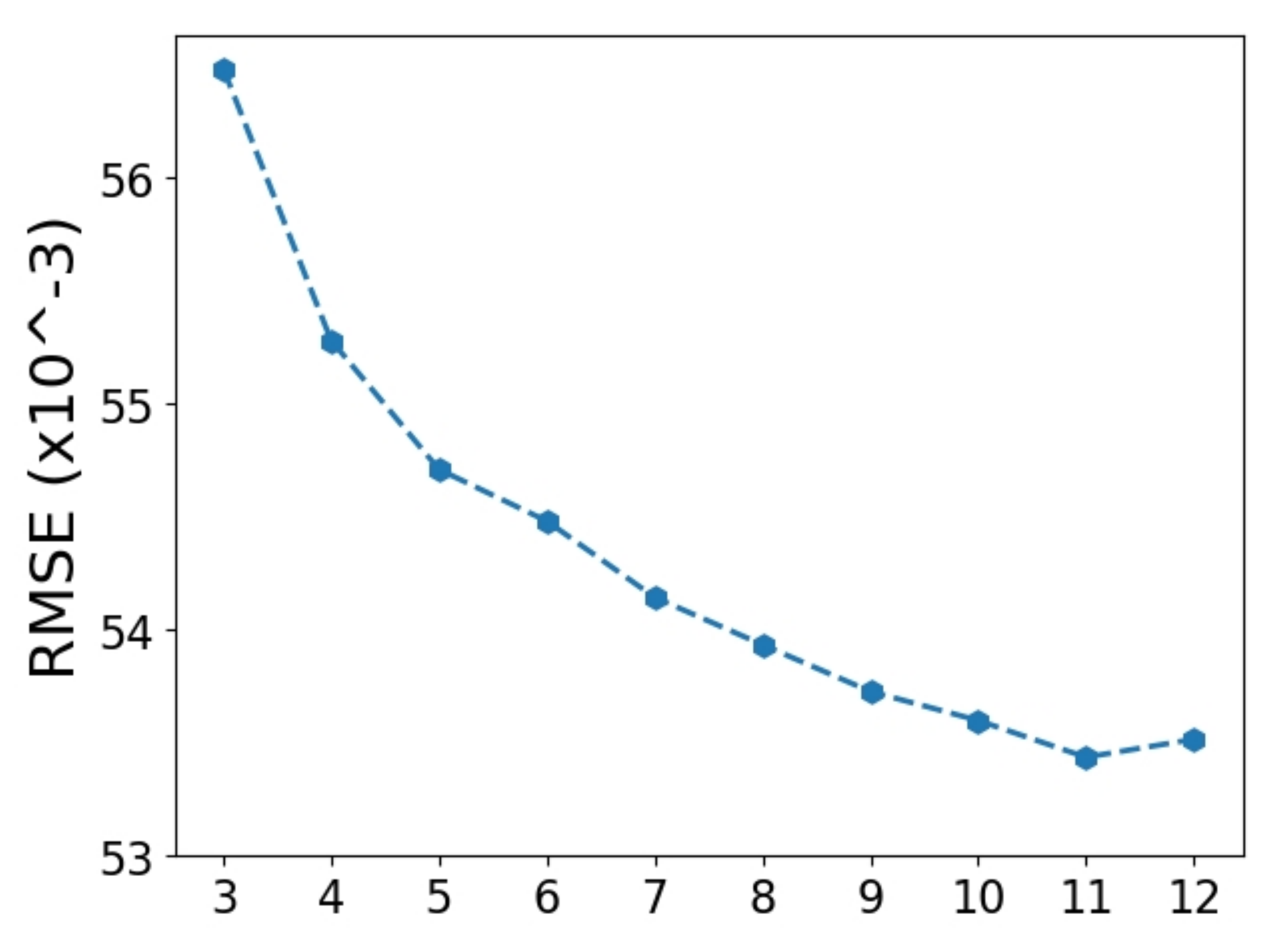}
    \label{fig:depth}
}\subfigure[Minimum number of samples to split an internal node (S)]{
    \includegraphics[width=0.31\textwidth]{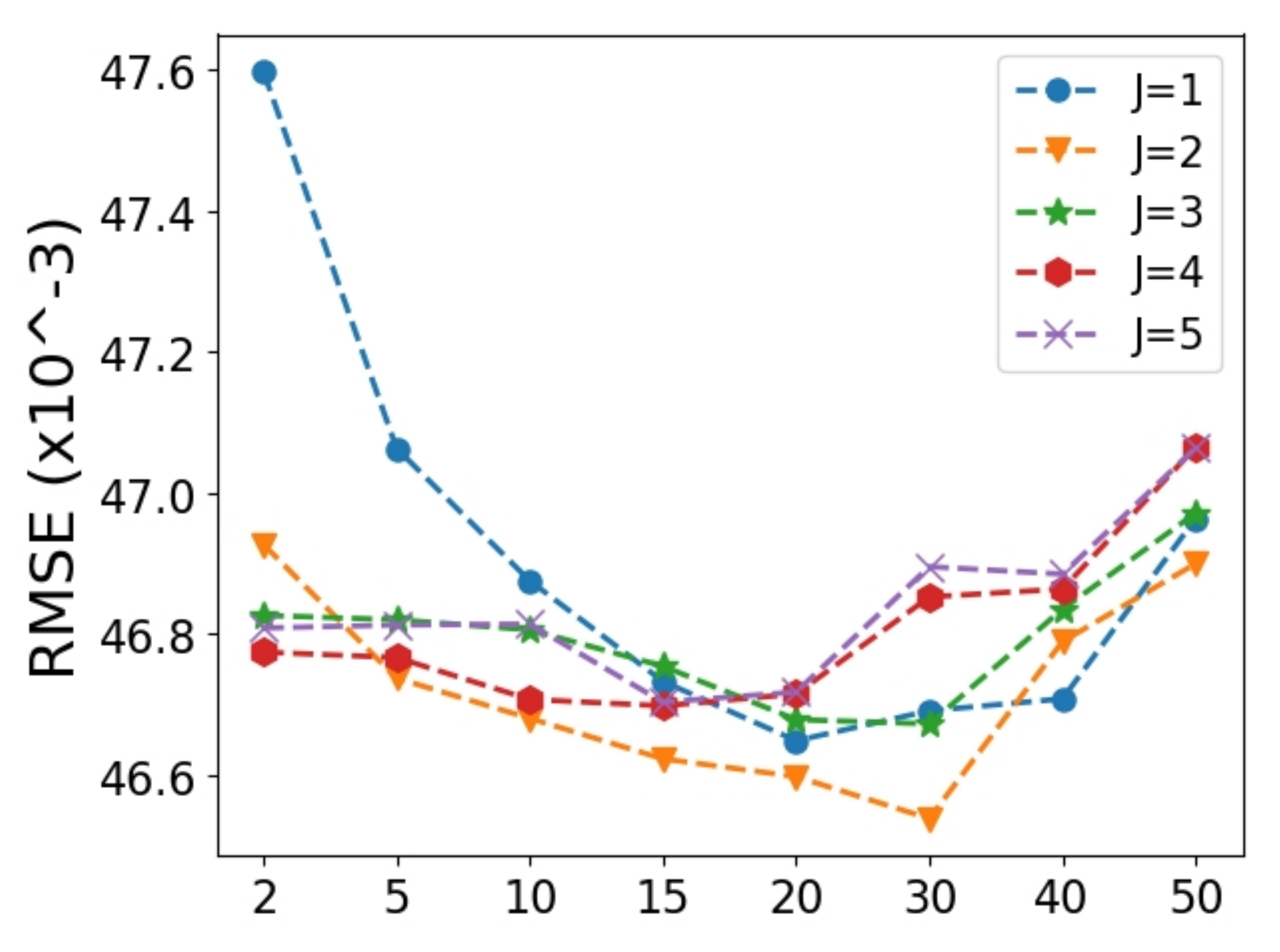}
    \label{fig:evalleaf-split}
}
\caption{Effect of various learning parameters on the accuracy of the prediction. Y-axis represents the error rate. X-axis of (a), (b), and (c) represent the number of trees in the GBDT algorithm, maximum depth of the decision tree, and the minimum number of samples to split a node (parameter S). Each line of (a) and (c) represent learning rate L and J values respectively. }
\end{figure*}

Let $r_{mi}$ denote the prediction error of record $i \in t$. Recall that the core idea of GBDT is to learn from and improve upon the mistakes of the previous iteration. Accordingly, on line 5, we calculate $r_{mi}$ of the model created in the previous iteration (\ie $B_{m-1}(x)$). 
The value of $r_{mi}$ is calculated based on Equation~\ref{eq:gbdgrad}. In this equation, $y_i$ is the ground truth (\ie actual saving in Table~\ref{table:3}) for the prediction made by $B_{m-1}(x_i)$. Also, $L(y_{i}, B_{m-1}(x_{i}))$ denotes the loss function and it is calculated as explained in~\cite{friedman2002stochastic}. 

\begin{equation}
    r_{mi} = -{\begin{bmatrix}
          \frac{\partial L(y_{i} , B_{m-1}(x_{i}))}{\partial B_{m-1}(x_{i})}
          \end{bmatrix}} 
          \label{eq:gbdgrad}
\end{equation}


On line 7, the decision tree is updated (called $B_m(x)$) based on the value of $r_{mi}$. 
On line 9, the ensemble of created decision trees form the merge-saving prediction model. Details of forming the ensemble can be found in \cite{friedman2002stochastic}. 

\section{Performance Evaluation of the Execution-Time Saving Predictor}
\label{sec:eval}
To maximize the prediction accuracy and efficiency, it is critical to determine the optimal combination of parameter values used in the GBDT model. As such, in this section, first, we examine various parameters that influence the accuracy of the prediction model.  
The best performance is achieved by deliberately selecting the fittest combination of these parameters. 
The predicted time-saving is primarily used for scheduling purposes where prediction errors can perturb the scheduler. As such, we consider Root Mean Square Error (RMSE) as the primary performance evaluation metric.

Once we optimally configure the proposed GBDT model, in the second part, we measure and analyze its prediction accuracy with respect to other methods that can alternatively employed to predict the merge-saving.

\subsection{Tuning the Learning Rate of the Predictor Method}
\label{subsec:evalL}
Gradient boosting predictors become robust when the model is sufficiently learned. However, over-fitting can occur, if they learn too fast with too little variation in the input. The learning rate ($L$) of the predictor indicates how fast it can learn at each iteration. This parameter is generally considered along with the number of trees (denoted as $M$) that is used to train the model. Parameter $M$ is also known as the iterations parameter, because each iteration generates one tree.  

In this part, our goal is to tune the predictor with the appropriate learning rate. For that purpose, we examine the RMSE metric when the learning rate $L$ changes in the range of [0.5 , 0.005]. Each learning rate is examined when number of trees varies in the range of [350 , 6,000].

Fig.~\ref{fig:estimators} demonstrates the relationship between RMSE and $M$ for different values of $L$.
We observe that when the number of trees is low (\ie short training), higher learning rates lead to a faster converge of the model. Therefore, the model achieves high accuracy in a lower number of iterations. However, the high learning rate can be susceptible to noise on the gradient that impacts the accuracy when leaned with a relative high number of tree.

We observe the maximum prediction accuracy for low learning rates and high number of trees. 
Increasing $M$ and decreasing $L$ make the model less susceptible to the noise, however, it make the model more complex and time consuming. 
Accordingly, to strike a balance between accuracy and the model complexity, we configure $M=350$ and $L=0.1$.


\subsection{Tuning the Value of Regression Estimator Maximum Depth } Maximum Depth ($D$) is a parameter that controls the number of decision trees allowed in the model. The optimal value of $D$ varies from one model to another, depending on the interaction of features within the training dataset and other training parameters. This parameter can be ignored when there are only few features. However, in our model, the optimal depth value should be limited based on the interplay of the input parameters. 

Fig.~\ref{fig:depth} shows the correlation between maximum depth of the tree in the range of [3, 12] in the horizontal axis and its corresponding error rate (RMSE). We notice that, as the value of $D$ increases, the prediction accuracy continues to increase until $D$ reaches 12 where we have an inflection point and we observe over-fitting. Therefore, we set $D=11$ as the appropriate value for the task merging prediction method.

\subsection{Tuning the Value of Minimum Samples to Create Internal- and Leaf-Node }
\label{subsec:evalSJ}
In this part, we evaluate the parameters that control the minimum sample to create a new internal node and the minimum sample to create a new leaf node ($S$ and $J$ parameters, respectively) and measure their impact on the accuracy of the prediction model.

The value of $J$ parameter correlates with the  value of $S$ parameter. Accordingly,
in Fig.~\ref{fig:evalleaf-split}, we explore the prediction accuracy (by means of the RMSE value in the vertical axis) obtained when the values of $S$ varies in the range of [2 , 50]. The experiment is conducted for different values of $J$ (in the range of [1 , 5]). 

We observe that regardless of the $J$ value, by increasing the value of $S$ a reverse bell curve shape is emerged. The lowest error rate, however, varies depending on the value of $J$ parameter. The rebound of error rate indicates overfitting and should be avoided. From this experiment, we configure $J=2$ and $S=30$ that offer the lowest error rate.


\subsection{Evaluating Improvement in the Prediction Accuracy}

In this part, we evaluate accuracy of the proposed prediction model (when configured as: \{ $M=350$, $L=0.1$, $D=11$, $S=30$, $J=2$ \}) against two alternative prediction methods. The first baseline approach, called \emph{Na\"ive} predictor, carries out the prediction based on a lookup table of mean execution-time saving for each operation. Another baseline approach is based on machine learning and uses a multi-layer perceptron (MLP)~\cite{plonis2020prediction} for prediction.

\begin{figure}[]
    \centering
    \includegraphics[width=0.31\textwidth]{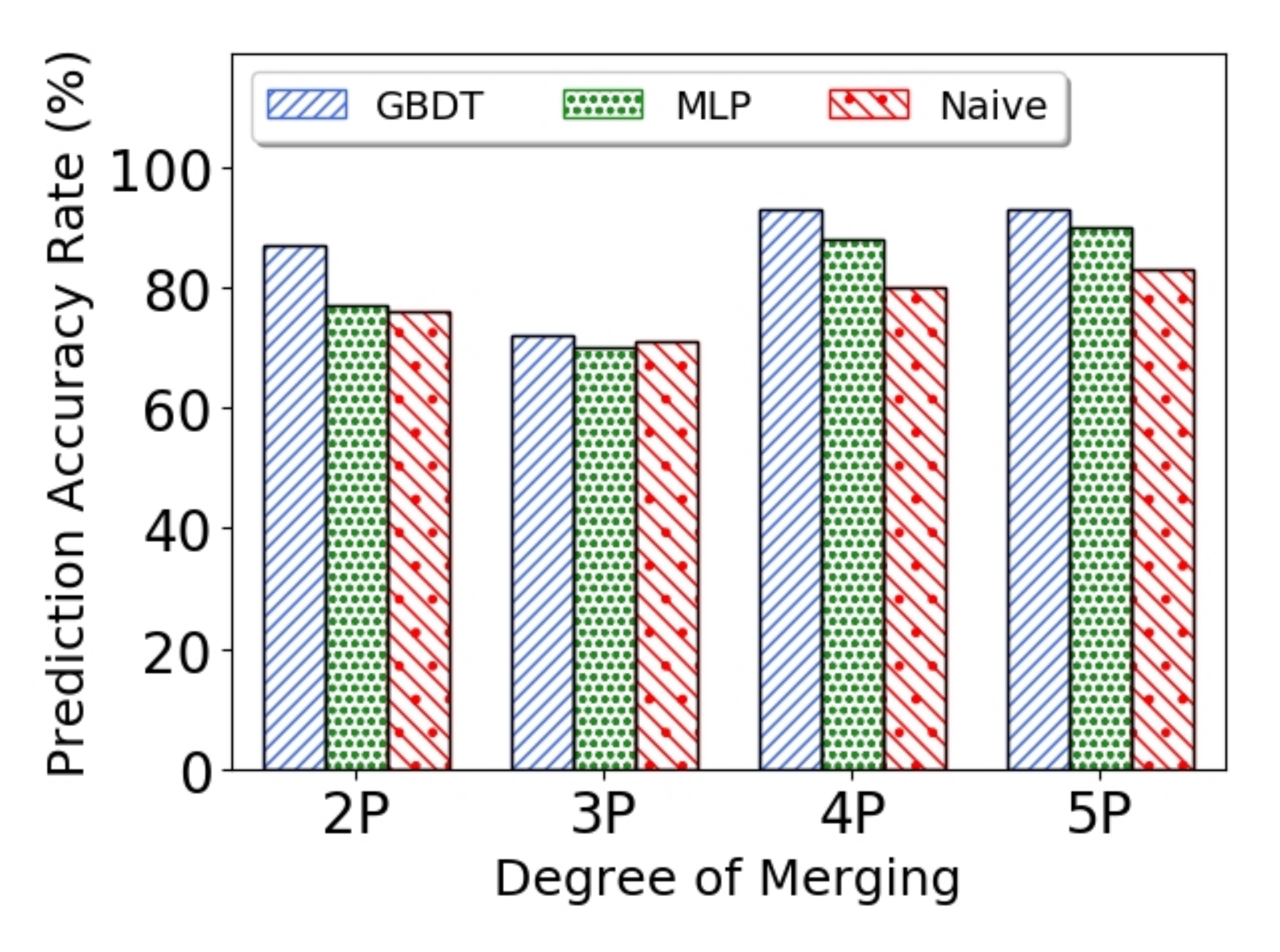}
    \caption{Comparing the prediction accuracy of proposed execution-time saving prediction model (GBDT) against MLP and Na\"ive approaches. The horizontal axis represents the number of tasks merged to create a merged task and vertical axis represents the percentage of cases accurately predicted.}
    \label{fig:comparison}
    \vspace{-10pt}
\end{figure}

The prediction accuracy is reported as the percentage of correct predictions, denoted as $C$ and is defined based on Equation \ref{eq:accuracy}. In this equation, $A$ represents the total number of test cases, $P$ is the predicted execution-time saving ratio, $E$ is the observed execution-time saving ratio, and $\tau$ is the acceptable error rate, which is set to  0.12 in Fig.~\ref{fig:comparison}.

\begin{equation}
    C = 100\% \times \frac{1}{A}\sum_{i=1}^{A}
    \begin{cases}
    0, & |P_i- E_i| > \tau \\
    1, & |P_i- E_i| \leq \tau
    \end{cases}
    \label{eq:accuracy}
\end{equation}

We observed that 
the GBDT model significantly outperforms the prediction accuracy of MLP and Na\"ive approaches, regardless of   merging degree. 
Both MLP and GBDT significantly perform more accurate for higher degrees of merging (4P and 5P) than the lower ones (2P and 3P). The reason is that, the lower degree of merging saves relatively low amount of execution-time, which is difficult to accurately predict. The maximum accuracy is 93\% when GBDT is employed in 4P.
\section{Conclusion and Future works}\label{sec:conclsn}
In this research, we studied the potential of reusing computation via merging similar tasks to reduce their overall execution-time in the clouds.
Considering video processing context, we built a video benchmarking dataset and evaluated the parameters that influence the merge-saving. We observed that merging similar video processing tasks can save up to 31\% (for merging two tasks) of the execution-time that implies a significant cost saving in the cloud. We also learned that the merge-saving gain becomes negligible, when degree of merging is greater than three.
Then, we leveraged the collected observations to train a machine learning method based on Gradient Boosting Decision Trees (GBDT) to predict the merge-saving of unforeseen task merging cases. The fine-tuned prediction model can provide up to 93\% 
accurate saving prediction.
The next step following this study is to explore an even broader variety of operations in other contexts. Rather than a single level predictor, a future work can utilize multi-level predictor where the first level predict the operation behavior, then the second level predict the merge-saving based on the parameters.

\section*{Acknowledgments}
This research was supported by the Louisiana Board of Regents under grant number LEQSF(2016-19)-RD-A-25. 

\bibliographystyle{IEEEtran}
\bibliography{references}
\end{document}